\documentclass[preprint,showpacs,preprintnumbers,amsmath,amssymb]{revtex4}

\usepackage{graphicx}
\usepackage{dcolumn}
\usepackage{bm}
\usepackage{fancybox}

\newcommand{\bfq}{{\bf q}}
\newcommand{\bfk}{{\bf k}}
\newcommand{\bfr}{{\bf r}}

\def\qsgw{QS{\em GW}}

\def\H0{H^0}

\newcommand{\req}[1]{\mbox{Eq.~(\ref{#1})}}

\def\rs{r_{\rm s}}

\def\iDelta{{\it \Delta}}
\def\efermi{\mbox{$E^{\rm F}$}}

\def\connect#1{\leavevmode{\setbox1=\hbox{#1}\copy1%
\raise .2\ht1 \vbox{\moveleft \wd1\vbox{\hrule width \wd1 height .5pt depth 0pt}}%
}}

\def\ftn[#1]{\rlap{\footnotemark[#1]}}

\def\rs{r_{\rm s}}
\def\efermiv{E^{\rm Fermi}_{\rm val}}
\def\efermic{E^{\rm Fermi}_{\rm con}}

\def\mv{m_{\rm val}}
\def\mc{m_{\rm con}}
\def\Nc{N_{\rm con}}
\def\Nv{N_{\rm val}}

\def\kf{{\bf k}^{\rm F}}
\def\sigx{\Sigma^{\rm x}}
\def\sigc{\Sigma^{\rm c}}
\def\eh{\epsilon^{\rm Hartree}}
\def\ek{\epsilon^{\rm kin}}

\def\iDelta{{\it \Delta}}
\def\eg{E_{\rm g}}
\def\egc{E_{\rm g}^{\rm cr}}
\def\eov{E_{\rm ovl}^{\rm cr}}
\def\eps{\bar{\varepsilon}}
\begin{document}

\title{First-order metal-insulator transition in band overlap mechanism}
\author{Takao Kotani}
\affiliation{School of Materials, Arizona State University, Tempe, AZ, 85284}

\author{Rei Sakuma}
\affiliation{Graduate School of Advanced Integration Science, Chiba University, Chiba 263-8522, Japan}

\date{\today}

\date{\today}

\begin{abstract}
We present a method to analyze the metal-insulator transition (MIT)
due to the band overlap mechanism. It is based on
a model with the knowledge of the homogeneous electron gas,
combined with results based on the quasiparticle self-consistent 
$GW$ method. Because of the long-range nature of the Coulomb interaction, 
the MIT occurs as the first-order phase transition, 
that is, the band gap becomes negative (band overlap) suddenly 
at some critical lattice constant.
\end{abstract}


\maketitle
\begin{figure}[htbp]
\centering
\includegraphics[angle=0,scale=1]{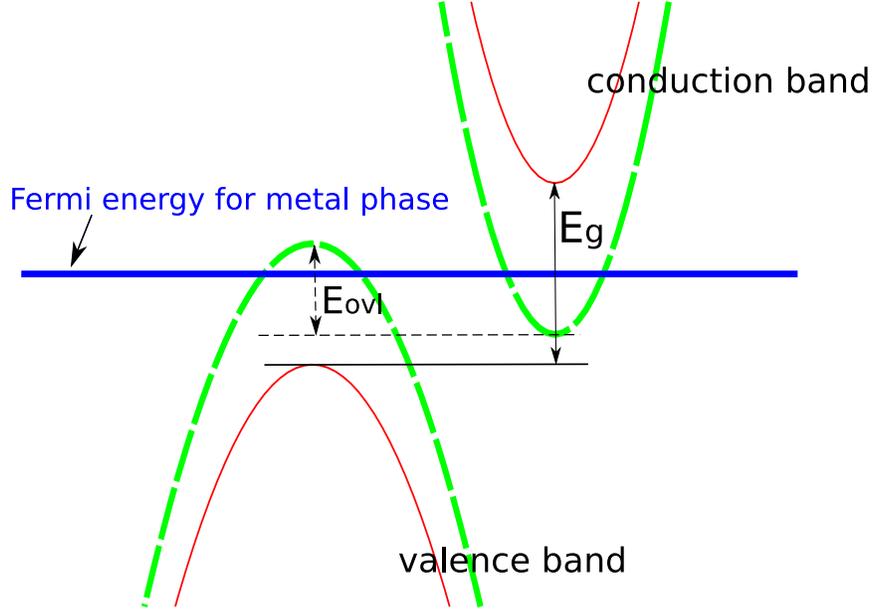}
\caption[]{(color online) 
Illustration of the first-order metal-insulator 
transition by the band overlap mechanism.
The band gap $\eg$ is gradually being reduced when
the lattice constant $a$ is getting smaller. 
Then it suddenly jumps to the metallic phase; the band overlap $E_{\rm ovl}$
appears as shown by the broken line. 
In our method explained in Fig.\ref{fig:fig2} around (see text), 
we assume the rigid band shifts. 
}
\label{fig:mit}
\end{figure}
The metal-insulator transitions (MIT) is important not only
for fundamental physics, but also for
the high-pressure physics, or for its potential 
applicability to electrical or optical switches, e.g, YH3 \cite{Huiberts96}.
Among the possible mechanisms of the MIT \cite{imada98},
we focus on the band-overlap mechanism  
which is the simplest in the sense that it is described 
within the one-particle picture. 
Here we present a theoretical treatment at
zero temperature without phonons.
The MIT is illustrated in Fig.\ref{fig:mit} (we explain it below).
Though the density functional theory (DFT)
can give an one-particle picture represented by 
its Kohn-Sham eigenvalues and eigenfunctions, 
the eigenvalues can not be identified as the quasiparticle energies (QPEs); 
it is well known that the DFT predicts too small $\eg$. 
The problem is not in the local density approximation 
(LDA) usually used in DFT;
as shown by the calculations with the optimized effective potential method 
with the exact exchange plus correlation in the random phase approximation 
(EXX+RPA), Kotani showed that the true Kohn-Sham eigenvalues without LDA 
are only slightly larger than the eigenvalues in LDA for semiconductors 
\cite{kotani98}. It is confirmed by other groups recently \cite{gruning06}.

Thus it is necessary to use a method beyond DFT to obtain
the reasonable quasiparticle (QP) picture, or rather
the one-particle static Hamiltonian $\H0$ which represent the QP.
For example, for MIT for bcc Hydrogen,
$GW$ calculations is used by Kioupakis, Zhang,Cohen, and Louie 
\cite{kioupakis08}. The $GW$ approximation ($GW$A) \cite{ferdi98} 
is theoretically reasonable to obtain QPEs, 
however, the reliability of the usual $GW$A is limited because it is 
the one-shot (perturbative) calculations 
starting from the solution in LDA. This can cause a problem in cases; 
for example, we may get insulator solution in $GW$A starting from the metallic 
solution in LDA. In such a case, its reliability is questionable. 
This situation can be corrected by the self-consistent perturbation idea.
In fact, they used a simplified version of the self-consistency
to determine $U$ in the LDA+$U$+$GW$ calculation \cite{kioupakis08} as was done 
by Aryasetiawan and  Gunnarsson \cite{ferdi94}. It is also used recently 
for the MIT problem for VO$_2$ by Sakuma Miyake and Aryasetiawan \cite{sakuma06}.
However, such a self-consistency is only for a degree of freedom;
a satisfactory version of self-consistency is formulated by 
the quasiparticle self-consistent $GW$ method (\qsgw).
Then all the degree of non-local static potential is 
determined self-consistently. We had shown that \qsgw\ reproduce 
the band gap $\eg$ for wide range of semiconductors and insulators 
very well \cite{mark06qsgw,kotani07qsgw}. 
Thus it is reasonable to apply \qsgw\ to analyze the MIT 
for the band-overlap mechanism.

The \qsgw\ is taken as an approximation 
to the rigorous theory. To see it, note 
that the self-consistent procedure in \qsgw\ 
\cite{kotani07qsgw,mark06qsgw,Faleev04}
can be divided into two parts. 
One is the determination of the one-particle static Hamiltonian $\H0$ 
from the given Hartree potential plus self-energy $V_{\rm H} +\Sigma(\bfr, \bfr',\omega)$,
the other is the perturbative calculation of 
$V_{\rm H} +\Sigma(\bfr, \bfr',\omega)$ in $GW$A
starting from $\H0$. The former is a recipe
to extract $\H0$ which represent the QP;
note that iterative procedure is required to determine $\H0$
even for fixed $V_{\rm H} +\Sigma(\bfr, \bfr',\omega)$ 
(see the norm minimization formalism \cite{mark06qsgw}).
The latter can be replaced by the rigorous procedure 
at least as a thought experiment. If we were able to do it,
we would have reached to the exact ground state with reasonable $\H0$.
A practical method to improve \qsgw\  is presented
by Shishkin, Marsman and Kresse \cite{shishkin07}, 
where the excitonic effects (correlational motion
of the electron-hole) are included in the polarization function 
when we evaluate the screened Coulomb interaction $W$. 
They succeeded to give systematic improvement 
for the overestimation of $\eg$ in the \qsgw.

Here we present a method to analyze the MIT
based on the knowledge of homogeneous electron gas
in combination with the insulator solutions 
given by \qsgw\ (or its extension).
As seen in Fig.\ref{fig:mit}, we have valence and conduction 
bands with $\eg$ for the insulator phase.
$\eg$ is monotonically getting smaller when 
the lattice constant $a$ is being reduced 
by external pressure. If we have a first-order MIT,
$\eg$ suddenly jumps from some positive value 
to the negative value (band overlap)
at the critical lattice constant $a=a_{\rm c}$.
This occurs if the total energy for the metal 
phase becomes lower than that of the insulator.
Then we see electron pocket in valence band
and hole pocket in conduction band, which are 
specified by the Fermi energy $\efermi$.
Our analysis below shows that the first-order MIT is inevitably occur 
because of the nature of the long-range Coulomb interaction. 
Then we estimate the size of the transition for the case of fcc YH3 
based on the \qsgw\ calculation for insulator phase.

Before looking into our model, let us remind the nature of
the homogeneous electron gas \cite{hedin65,fetter71}. 
At lower density $n$ (density $n$ can be specified by $\rs$, 
$4 \pi\rs^3/3  =1/n$),  the Coulomb interaction $v$ is more important 
than the kinetic energy.
The exchange self-energy at $\efermi$ as
$\sigx= -\frac{4}{3} \frac{0.916}{\rs}=-1.2218/\rs$ Ry 
dominates the kinetic energy $\ek=|\kf|^2/(2m)=3.6832/\rs^2$ Ry at 
low density (large $\rs$),
where $\kf$ is the Fermi momentum.
The Fermi energy is given as
$\efermi= \ek + \sigx + \sigc$, where the Hartree term do not exist 
since it is cancelled by the background positive charge.
These terms are evaluated at $\efermi$ and at $\kf$ for given $n$.
Because of the behavior of $\ek$ and $\sigx$, 
$\efermi$ as function of $n$ is not monotonic.
$\sigx(n)\propto - n^{1/3}$ dominates $\efermi$ for $n \to 0$,
though $\ek \propto n^{2/3}$ does $\efermi$ for high $n$.
$\sigc$ enhances the effect of $\sigx$
as seen in Table III in Ref.\onlinecite{hedin65} by Hedin,
where we see that $\efermi$ is negative at $\rs \gtrsim 3$.
This large negative values of $\sigx + \sigc$ 
overriding $\ek$ at low $n$ means the energy gain due to the Fermi 
statistics and the correlational motion of electrons 
(or the structure of the full many-body eigenfunctions).
Since the Coulomb interaction is stronger at $\bfq \to 0$, 
the behavior of $\efermi$ is quite anomalous at $n \to 0$.
This is in contrast to models with short-range interaction,
which is $q$-independent and we have $\sigx \propto - n$.
Then it can not dominate the kinetic term at low density.

We treat a simplified model to avoid difficulty in real systems. 
In principle, we treat both of the insulator and the metal phases 
within \qsgw (or its extensions).
However, in practice, simple sampling method for the Brillouin-zone 
summation \cite{kotani07qsgw} is not applicable 
to the metal phase because we need to use 
too many $\bfk$ points to take into account the contributions 
from the small amount of holes in valence band (and electrons in conduction bands).
To avoid this problem, we consider a model with simplified energy 
bands with some assumptions. At first, we have to prepare $\H0$ for 
the insulator phase by \qsgw. When we can observe the MIT, 
$\H0$ should show small $\eg$, which is decreasing as $a$ is being reduced. 
In our model, we only take into account 
the bands around the Fermi energy as shown in Fig.\ref{fig:mit}.
The valence bands are specified by the isotropic effective mass $\mv$ and
its degeneracy $\Nv$; the conduction band by $\mc$ and $\Nc$. 
These parameters should be chosen so as to mimic the bands in 
the insulator phase given by \qsgw.
As we concentrate on the MIT, we only take into account 
the monotonic $a$-dependence in $\eg$; 
we assume $\mc$ and $\mv$ are not $a$-dependent.
Further, we assume the rigid shift of energy bands 
(no deformations). 

For the model, we consider the situation
that some amount of electrons (specified by density $n$) 
are moved from the valence bands to the conduction bands.
Then we treat two quasi-Fermi energies, $\efermiv(n)$ 
for the electrons in valence bands, 
and $\efermic(n)$ for holes in conduction bands.
As we discuss below, we will evaluate the total energy 
as function of $n$, and the energy minimum of the model 
occurs at some finite $n$ (thus metal) below some critical $\eg$.

$\efermiv(n)$ and $\efermic(n)$ 
are defined as the changing rate of the total energy per adding 
(subtracting) an electron. 
Thus the total energy $\iDelta E(n)$ relative to the insulator 
phase is given by an adiabatic connection as
\begin{eqnarray}
\iDelta E(n) = \int_0^{n} dn (\eg+\efermic(n)-\efermiv(n) ),
\label{deltae}
\end{eqnarray}
where we can set $\efermiv(0)=\efermic(0)=0$. 
Then $\efermic(n)$ is given as
$\efermic(n)= \ek(n) + \eh(n)+ \sigx(n) +\sigc(n)$. 
As we saw in the electron gas, 
$\ek(n)= |\kf|^2/(2\mv) \propto \frac{n^{2/3}}{\mv} $. 
The exchange part is given as $\sigx=-{1.2218}/({\eps \rs})$,
where $\eps$ means the effective dielectric constant representing the 
screening effect in the insulator phase, 
and $\rs$ is for the density $n$.
We have to use $n/\Nv$ instead of
$n$ if $\Nv\ne 1$. We treat $\eps$ as constant;
$\eps$ is little dependent on $a$ in the case of indirect gap as in fcc YH3.
For $\sigc$, we use the RPA formula, e.g, see Eq.(86) in Ref.\cite{hedin65},
where we use $W= 
v/\varepsilon(\bfq, \omega)= v/{\eps} \times 1/(1-v\chi_0/\eps)$.
This $W$ is obtained in the RPA with
$v/\eps$ instead of the bare Coulomb interaction $v$. 
$\chi_0$ here contains the contribution due to electrons in valence bands
and due to holes in conduction bands, but no interband contributions;
neglecting the interband contribution will be reasonable
for the case with indirect gap.
We neglect $\eh$, because $\eh \propto n$
can be neglected in comparison with $\sigx$ at least for small $n$. 
We further neglect the other kind of correlational effect beyond \qsgw\
as excitonic effects between electrons. 
With these assumptions, we can evaluate
$\efermic(n)$ and also $\efermiv(n)$ 
for given $\mv,\Nv,\mc,\Nc$, and $\eps$.
In our model, $a$-dependence is only in $\eg$;
the dependence cause just a constant shift  
in the integrand of \req{deltae} as function of $n$. 
 
\begin{figure}[htbp]
\centering
\includegraphics[angle=0,scale=1]{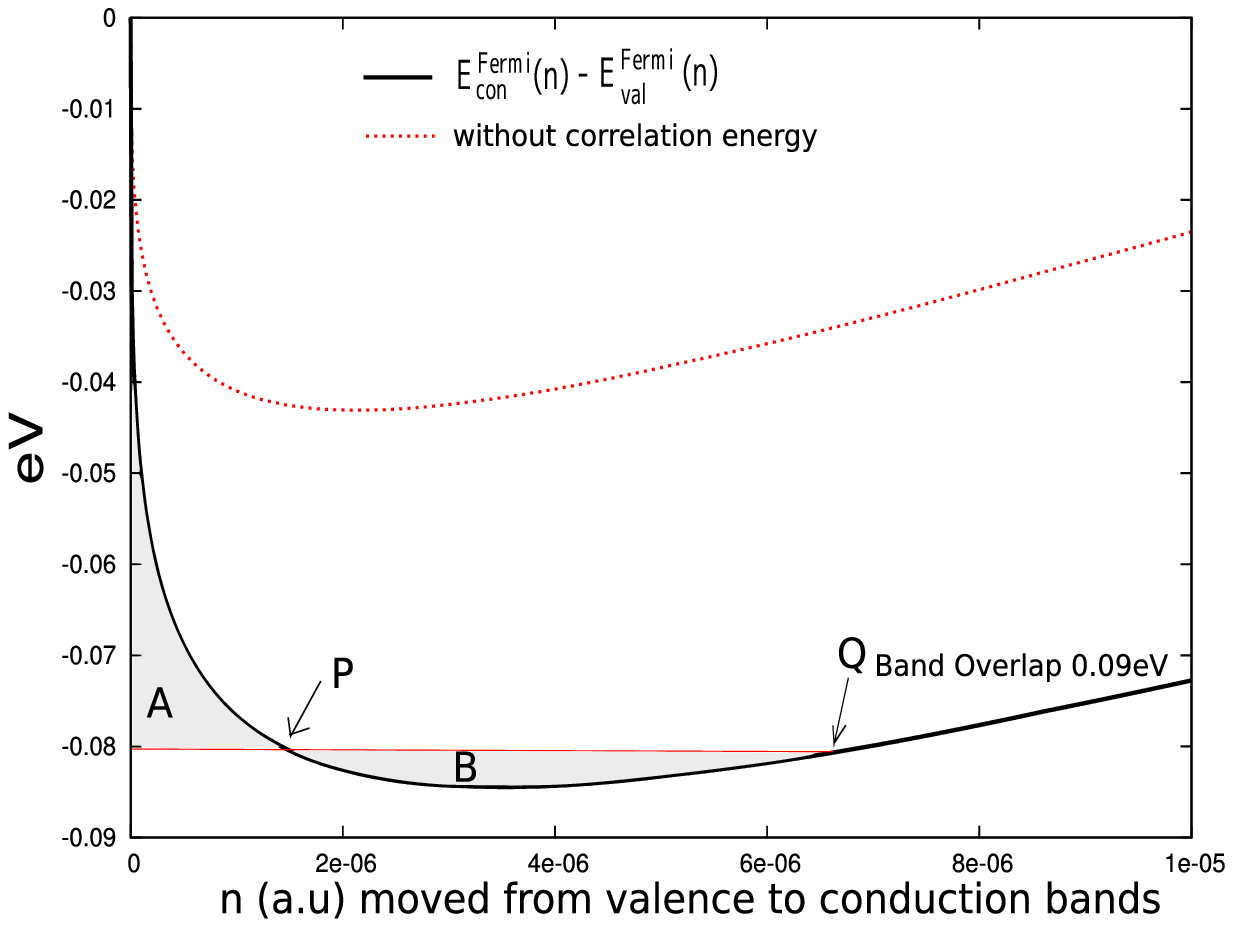}
\caption[]{Solid line shows $\efermic(n)-\efermiv(n)$ 
for the case $\mv=\mc=1$(in unit of electron mass), 
$\Nv=\Nc=1$, and $\eps=8$. Dotted line is without $\sigc$
contribution. 
The difference of areas A$-$B shows the total energy difference
between insulator and metal. When $\eg<0.08$ eV, we have $B>A$
resulting the stability of the metal phase at Q. See text.}
\label{fig:fig2}
\end{figure}

Let us look into the numerical results. 
In Fig.\ref{fig:fig2}, solid line shows $\efermic(n)-\efermiv(n)$
for the case $\mv=\mc=1$ (in unit of electron mass), 
$\Nv=\Nc=1$, and $\eps=8$. Because of the electron-hole symmetry 
in this case, $\efermic(n)-\efermiv(n) = 2 \efermic(n)$ is satisfied.
Derivative of the line is divergent at $n \to 0$ 
since it is $\propto -n^{1/3}$. At high density,
its behavior is controlled by $\ek \propto n^{2/3}$.
The integrand $\eg+\efermic(n)-\efermiv(n)$
in \req{deltae} is shown by the solid line ($-\eg$ as zero level). 
We show the case with $\eg=0.08$eV.
At the crossings P and Q, we have $\eg+\efermic(n)-\efermiv(n)=0$,
that is, the derivative $\frac{d \iDelta E}{d n} = 0$. 
$Q$ corresponds to the stable solution satisfying 
$\frac{d^2 \iDelta E}{d n^2} > 0$. 
The energy $\iDelta E$ of the metal phase at Q is given as the difference
of areas A$-$B. We have A$-$B=0 at $\eg=0.08$eV
in the case; for $\eg < \egc=0.08$eV,  the metal phase at Q has lower 
energy than the insulator phase since A$<$B. 
At Q point the band overlap is $\eov=$0.09 eV,
where we simply assume the rigid band shift  (no band narrowing nor widening). 
The dotted line is without $\sigc$; we see that
the contribution from $\sigc$ enhance $\egc$ almost twice larger.
Considering the behavior of $\efermic(n)-\efermiv(n)$,
this first-order phase transition can be a general 
phenomena for the band-overlap MIT.

\begin{table}
\caption{
\baselineskip 12pt
Calculated size of MIT in our model.
for given $\eps$, $\mv$, $\mc$, $\Nc$, (we use $\Nv=1$). 
When $\eg$ is getting smaller and reach at $\eg=\egc$, 
it cause the first-order phase transition 
to metal with the band overlap $\eov$.}
\begin{tabular}{c*6{@{\hspace*{0.5em}}r}}
\vbox{\vskip 2pt}
 $\eps$ & $\mv$ & $\mc$ & $\Nc$ & $\egc$ (eV) & $\eov$ (eV) \\
\colrule
 8 & 1 & 1   & 1    & 0.08  & -0.09  \\ 
 8 & .5 & .5 & 1    & 0.04  & -0.05  \\ 
 8 & 1 & 1   & 4    & 0.10  & -0.11  \\ 
 8 & .5 & .5 & 4    & 0.05  & -0.05  \\ 
 4 & .5 & .5 & 1    & 0.16  & -0.19  \\ 
 4 & 1 & 1  & 1     & 0.32  & -0.38  \\ 
 4 & .5 & .5 & 4    & 0.19  & -0.22  \\ 
 4 & 1 & 1  & 4     & 0.39  & -0.44  \\
 8 & .41 & .92 & 4  & 0.05  & -0.06  \\ 
 4 & .41 & .92 & 4  & 0.21  & -0.24  \\ 
\end{tabular}
\label{tab:mitsize}
\end{table}

In Table \ref{tab:mitsize}, calculated $\egc$ and $\eov$ are given
for kinds $\mv,\mc,\Nv$ and $\eps$. 
As $\mv$ and $\mc$ are heavier, $\egc$ and $\eov$ 
get larger because the contribution from kinetic term 
is smaller. $\eps$, which determines the size of the effective 
Coulomb interaction, can strongly affect on $\egc$ and $\eov$.
We may overestimate $\sigc$ a little in RPA, thus we does $\egc$; 
this is indicated by the fact that $\sigc$
for homogeneous electron gas given by the
accurate method \cite{vwn} is 69 \% of RPA value 
at $\rs$=4 (62 \% at $\rs=$100). 
The parameter set $\mv=.41,\mc=.92,\Nc=4$,and $\eps=8$ 
is to mimic the energy bands of fcc YH3.
Its bottom of conduction band is at $L$ point, where 
two Fermi surface exist, but we neglect the smaller one
because it contains only $\sim$10\% of electrons of bigger Fermi surface.
The Fermi surface is anisotropic, but we take 
the simple average as $\mc=(m_xm_ym_x)^{1/3}=0.92$.
The valence top is at $\Gamma$, where it has one spherical 
Fermi surface. We will report how the band gap changes as
the function of $a$ elsewhere together with other analysis 
for comparison with experiments \cite{sakumayh3}.
In anyway, our result of $\egc=0.05$eV should be taken as
a semi-quantitative prediction since out model treatment
is very simplified.

Though our MIT mechanism may occur in reality,
we have not found experiments which are directly related 
to our MIT mechanism. As we discuss elsewhere\cite{sakumayh3}, 
the MIT for YH3 observed in experiments will be
mainly controlled by the structural transition,
thus the band-overlap mechanism here will be not 
directly related to the experiments.
In the case of GdN \cite{chantis07f}, the MIT explained here might occur.
See Fig.2 in Ref.\onlinecite{chantis07f}.
However, we need to improve our treatment to have 
some numerical prediction for GdN 
because it is magnetic and multiple bands are involved.
As for the bcc hydrogen \cite{kioupakis08}, we may apply our MIT
mechanism. In fact, the variational Monte-Carlo results
seems to indicate the weak first-order phase transition
\cite{pfrommer98,zhu90}.


In conclusion, we have given a theoretical analysis 
for the metal-insulator transition through the band-overlap mechanism.
We have showed that it occurs as the first-order phase transition;
when the band gap is getting smaller, the insulator phase
suddenly changes to the metal phase because of the energy gain 
due to the Fermi statistics and due to the correlational 
motion for the Coulomb interaction.
In our treatment, the self-consistency in \qsgw\
is not fully included; we only consider the rigid shift of energy bands. 
If we fully include the self-consistency, we should
have some deformation of the energy bands 
(changes of effective masses). However, we expect that our conclusion
will be unchanges qualitatively.
To treat kinds of materials in practice,
it will be necessary to improve our method.
Further, we will have to examine roles of other effects 
like excitonic effects or phonon effects.

This work was supported by ONR contract N00014-7-1-0479. 
We are also indebted to the Ira A. Fulton High Performance 
Computing Initiative.

\bibliography{lmto}
\end{document}